\documentclass[%
aip,jap,
 amsmath,amssymb,floatfix,
 preprint,%
]{revtex4-1}

\usepackage[utf8]{inputenc}
\usepackage[T1]{fontenc}
\usepackage{array}
\usepackage{amsmath,amssymb,amsfonts}
\usepackage{algorithmicx}
\usepackage{mathtools}
\usepackage{algorithm}
\usepackage{algpseudocode}
\usepackage{graphicx}
\usepackage{multirow}
\usepackage{textcomp}
\usepackage{xcolor}
\usepackage{siunitx}
\usepackage{mathtools}
\DeclarePairedDelimiter\abs{\lvert}{\rvert}%

\def\BibTeX{{\rm B\kern-.05em{\sc i\kern-.025em b}\kern-.08em
    T\kern-.1667em\lower.7ex\hbox{E}\kern-.125emX}}

\begin{document}

\title{Would Magnonic Circuits Outperform CMOS Counterparts?}

\author{Abdulqader Mahmoud}
\email{a.n.n.mahmoud@tudelft.nl}
\affiliation{Delft University of Technology, 2628 CD Delft, The Netherlands}

\author{Nicoleta Cucu-Laurenciu}
\affiliation{Delft University of Technology, 2628 CD Delft, The Netherlands}
  
\author{Frederic Vanderveken}
\affiliation{Imec, 3001 Leuven, Belgium}

\author{Florin Ciubotaru}
\affiliation{Imec, 3001 Leuven, Belgium}

\author{Christoph Adelmann}
\affiliation{Imec, 3001 Leuven, Belgium}

\author{Sorin Cotofana}
\affiliation{Delft University of Technology, 2628 CD Delft, The Netherlands}

\author{Said Hamdioui}
\affiliation{Delft University of Technology, 2628 CD Delft, The Netherlands}

\begin{abstract}
\vspace{-0.6cm}
  In the early stages of a novel technology development, it is difficult to provide a comprehensive assessment of its potential capabilities and impact.  Nevertheless, some preliminary estimates can be drawn and are certainly of great interest and in this paper we follow this line of reasoning within the framework of the Spin Wave (SW) based computing paradigm. In particular, we are interested in assessing the technological development horizon that needs to be reached in order to unleash the full SW paradigm potential such that SW circuits can outperform CMOS counterparts in terms of energy consumption. In view of the zero power SWs propagation through ferromagnetic waveguides, the overall  SW circuit power consumption is determined by the one associated to SWs generation and sensing by means of transducers. While current antenna based transducers are clearly power hungry recent developments indicate that magneto-electric  (ME) cells have a great potential for ultra-low power SW generation and sensing. Given that MEs have been only proposed at the conceptual level and no actual experimental demonstration has been reported we cannot evaluate the impact of their utilization on the SW circuit energy consumption. However, we can perform a reverse engineering alike analysis to determine ME delay and power consumption upper bounds that can place SW circuits in the leading position. To this end, we utilize a $32$-bit Brent-Kung Adder (BKA) as discussion vehicle and compute the maximum ME delay and power consumption that could potentially enable a SW implementation able to outperform its \SI{7}{nm} CMOS counterpart. We evaluate different BKA SW implementations that rely on conversion- or normalization-based gate cascading and consider continuous or pulsed SW generation scenarios. Our evaluations indicate that \SI{31}{nW} is the maximum transducer power consumption for which a $32$-bit Brent-Kung SW implementation can outperform its \SI{7}{nm} CMOS counterpart in terms of energy consumption.
\end{abstract}

\maketitle

\section{Introduction}
In the last decade, there was an enormous increase in the raw data generation and collection because of information technology ubiquity. In addition, the processing and analysis of these huge amounts of data requires high performance computing platforms \cite{data1}, which, during the last decades, where enabled by the CMOS downscaling iduced performance improvement. However, Dennard scaling becomes  more and more difficult due to leakage, reliability, and economical walls \cite{cmosscaling1},which predicts the near end of Moore's law. As a result, researchers have investigated alternative technologies such as graphene devices \cite{Yande1}, memristors \cite{memristor10}, and spintronics \cite{ITRS}. One of the spintronic technologies is the Spin Wave (SW) technology, the so called magnonics, which is promising for computing as \cite{amahmoud2,amahmoud1,parallelism,fanout10}: i) it has ultra-low energy consumption as the electrons are spinning and not moving, ii) it is highly scalable as the SW wavelength, which is the only scalability limitation, can reach down to few nanometers, iii) it has an acceptable delay, and iv) it has natural support for parallelism feature as SWs with different frequencies can simultaneously propagate through the same waveguide without affecting each other.

Motivated by the SW interaction potential, researchers have investigated different SW logic gates and circuits \cite{logic21,fanout, parallelism,parallelism1, fanout10, fanout11, logic1,amahmoud1,SW_app_comp,SW_compressor,approx1,memory3}. In \cite{logic21}, a Mach-Zehnder interferometer has been utilized to design the first experimental SW NOT gate, whereas the Mach-Zehnder interferometer has been used to build different single output gates such as Majority, (N)AND, (N)OR, and X(N)OR gates in \cite{logic21}. On the other hand, multi-output logic gates have been reported in \cite{fanout, fanout10,fanout11}. Moreover, the unique parallelism feature of the SW technology has been utilized to demonstrate multi-frequency logic gates \cite{parallelism,parallelism1}. On the circuit side, different circuits have been illustrated at the conceptual level  \cite{logic1}, simulation level \cite{amahmoud1,SW_app_comp,SW_compressor,approx1}, and physical millimeter scale level \cite{memory3}. 
In \cite{survey2,survey3,survey4,Excitation_table_ref16}, benchmarking of beyond CMOS technologies, e.g., SW, Graphene, SpinFET, has been performed by modelling these devices as a capacitor and parasitics, which is not quite appropriate for the SW case. Thus, to get better inside on the SW based computation paradigm potential, which is certainly a topic of great interest, a  benchmarking based on more realistic device models and that considers the complications related to SW gate cascading and fanout achievement is required.

In view of the zero power SWs propagation through ferromagnetic waveguides, the overall  magnionic circuit power consumption is determined by the one associated to SWs generation and sensing by means of transducers. Recent developments indicate that magneto-electric  (ME) cells \cite{amahmoud2} have a great potential for ultra-low power SW generation and sensing, but they have been only proposed at the conceptual level and no actual experimental demonstration has been reported. 

\begin{figure}[t]
\centering
  \includegraphics[width=0.3\linewidth]{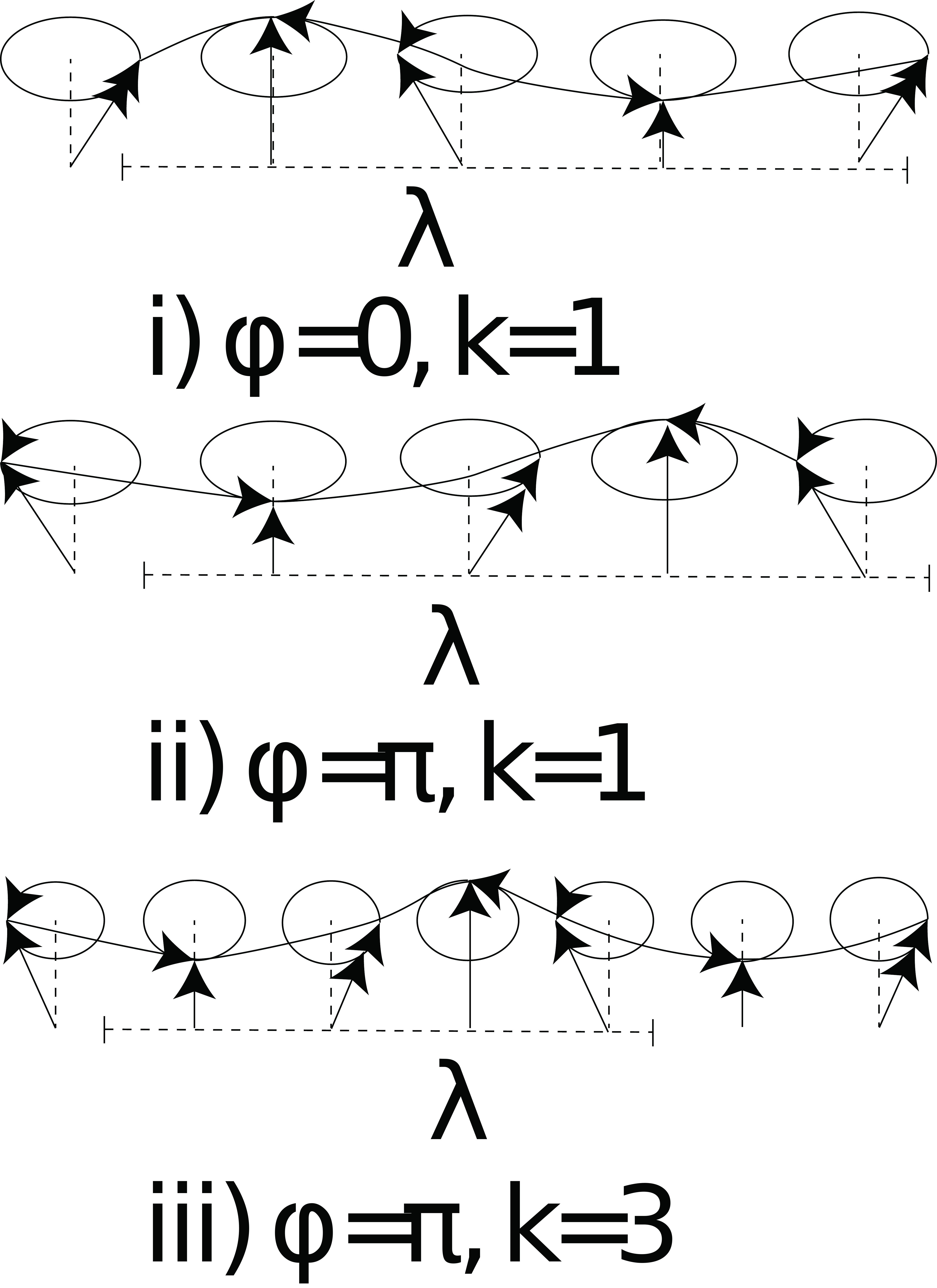}
  \caption{Spin Wave Parameters.}
  \label{fig:spin_wave_parameters}
  \vspace{-0.6cm}
\end{figure} 

Thus, since SW technology is still under development and the necessary insight for assessing its practical potential is not available, we reformulate in this paper the benchmarking process as a quest to identity the technological limits that need to be reached in order to fully unleash the SW potential such that SW circuits can outperform CMOS counterparts in terms of energy consumption. Thus, we assume a $32$-bit Brent-Kung Adder (BKA) as discussion vehicle, and perform a reverse engineering process in order to determine the transducer power upper bound that allows its SW implementation to outperform the \SI{7}{nm} CMOS counterpart. We evaluate different BKA SW implementations that rely on  conversion- or normalization-based gate cascading, while considering continuous or pulsed SW generation scenarios. Our evaluations indicate that \SI{31}{nW} is the maximum transducer power consumption for which a $32$-bit Brent-Kung SW implementation can still outperform its \SI{7}{nm} CMOS counterpart in terms of energy. Based on this assessment, we conclude the paper by discussing challenges and possible future directions toward building efficient magnonic circuits.

The paper organization is as follows. Section \ref{sec:Basics of spin-wave technology} explains the SW fundamentals and computing paradigm. Section \ref{sec:Transducer Power Upper Bound} provides the possible implementation of the $32$-bit Brent Kung Adder in SW technology. Section \ref{sec:Conclusion} sums-up the paper.

\section{Spin Wave Fundamental and Computing Paradigm}
\label{sec:Basics of spin-wave technology}

For out of equilibrium magnetization states, the magnetization dynamics is governed by the Landau-Lifshitz-Gilbert (LLG) equation \cite{amahmoud2}
\vspace{-0.2cm}
\begin{equation} \label{eq:1}
\frac{d\vec{M}}{dt} =-\abs{\gamma} \mu_0 \left (\vec{M} \times \vec{H}_{eff} \right ) + \frac{\alpha}{M_s} \left (\vec{M} \times \frac{d\vec{M}}{dt}\right ),
\end{equation}
where $\gamma$ is the gyromagnetic ratio, $\mu_0$ the vacuum permeability, $M$ the magnetization, $M_s$ the saturation magnetization, $\alpha$ the damping factor, and $H_{eff}$ the effective field, which consists of the external field, exchange field, demagnetizing field, and magneto-crystalline field.

For small magnetic disturbances, Equation (\ref{eq:1}) can be linearized and results in wave-like solutions, known as Spin Waves (SWs), which can be seen as the collective excitation of the electron spins in the magnetic material \cite{amahmoud2}. A SW  is described by its frequency $f$, amplitude $A$, phase $\phi$, wavelength $\lambda$, and wavenumber $k=\frac{2\pi}{\lambda}$ \cite{amahmoud2}, which are graphically presented in Figure \ref{fig:spin_wave_parameters}. The relation between the SW wavenumber and frequency is known as the dispersion relation and it is crucial for the SW circuit design \cite{amahmoud2}.

\begin{figure}[t]
\centering
  \includegraphics[width=\linewidth]{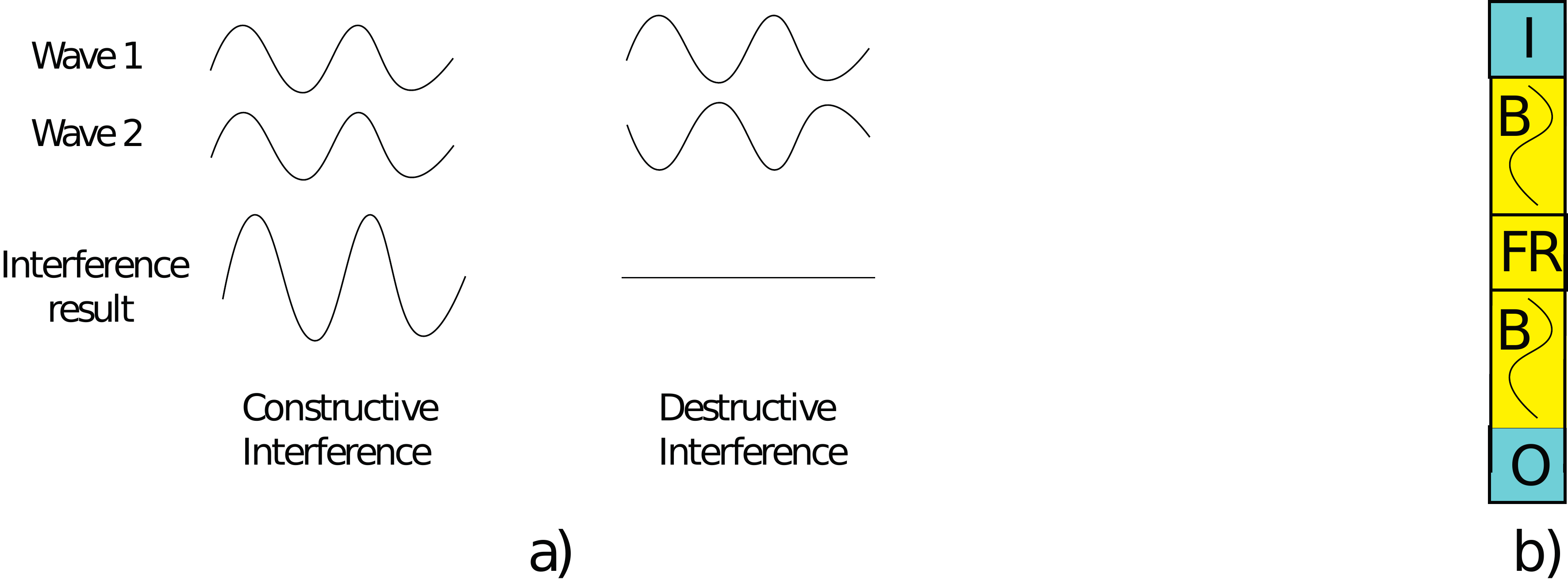}
  \caption{a) Constructive and Destructive Interference, b) Spin Wave Device.}
  \label{fig:spin_wave_device}
  \vspace{-0.6cm}
\end{figure}

SW amplitude, phase, and frequency can be utilized to encode information \cite{amahmoud2,parallelism}, which processing is governed by the wave interference principle. An example of SW interference is presented in Figure \ref{fig:spin_wave_device} a) where SWs are interfering constructively in the first case as they have the same phase $\Delta \phi=0$, and destructively in the second case as they are out-of-phase $\Delta \phi=\pi$. Moreover, assuming SW phase information encoding, i.e., phase $0$ and $\pi$ represents logic $0$ and $1$, respectively, SWs interaction supports Majority function evaluation. 
For instance, if $3$ SWs having the same amplitude, frequency, and wavelength interfere in the same waveguide, the resultant SW has $0$ phase, if at least $2$ SWs have $0$ phase, whereas the resultant SW has $\pi$ phase, if at least $2$ SWs have a $\pi$ phase. Note that such an implementation in CMOS requires $18$ transistors, whereas it can be directly implemented in SW technology with a single waveguide. One can easily deduce that more complex interference patterns can occur for SWs with different amplitude, frequency, wavenumber, and wavelength, which can be of great interest for developing future SW based  computing paradigms.

Figure \ref{fig:spin_wave_device} b) presents a general structure of the SW device, which consists of four main regions \cite{amahmoud2,Magnonic_crystals_for_data_processing}: i) Excitation region (I), ii) Waveguide (B), iii) Functional Region (FR), and iv) Detection region (O). At the excitation region, the SW is excited by means of voltage- or current- driven techniques such as microstrip antennas \cite{amahmoud2}, MagnetoElectric (ME) cells \cite{Fred20}, or spin orbit torques \cite{amahmoud2}. After the excitation, the SWs propagates through the magnetic waveguide and  reach the functional region, where it can be manipulated, i.e., amplified, normalized, or interfere with other SWs. Finally, at the detection region, the SW is detected by similar or different methods than in the excitation region  \cite{amahmoud2,Magnonic_crystals_for_data_processing}.

\begin{figure}[t]
\centering
  \includegraphics[width=0.9\linewidth]{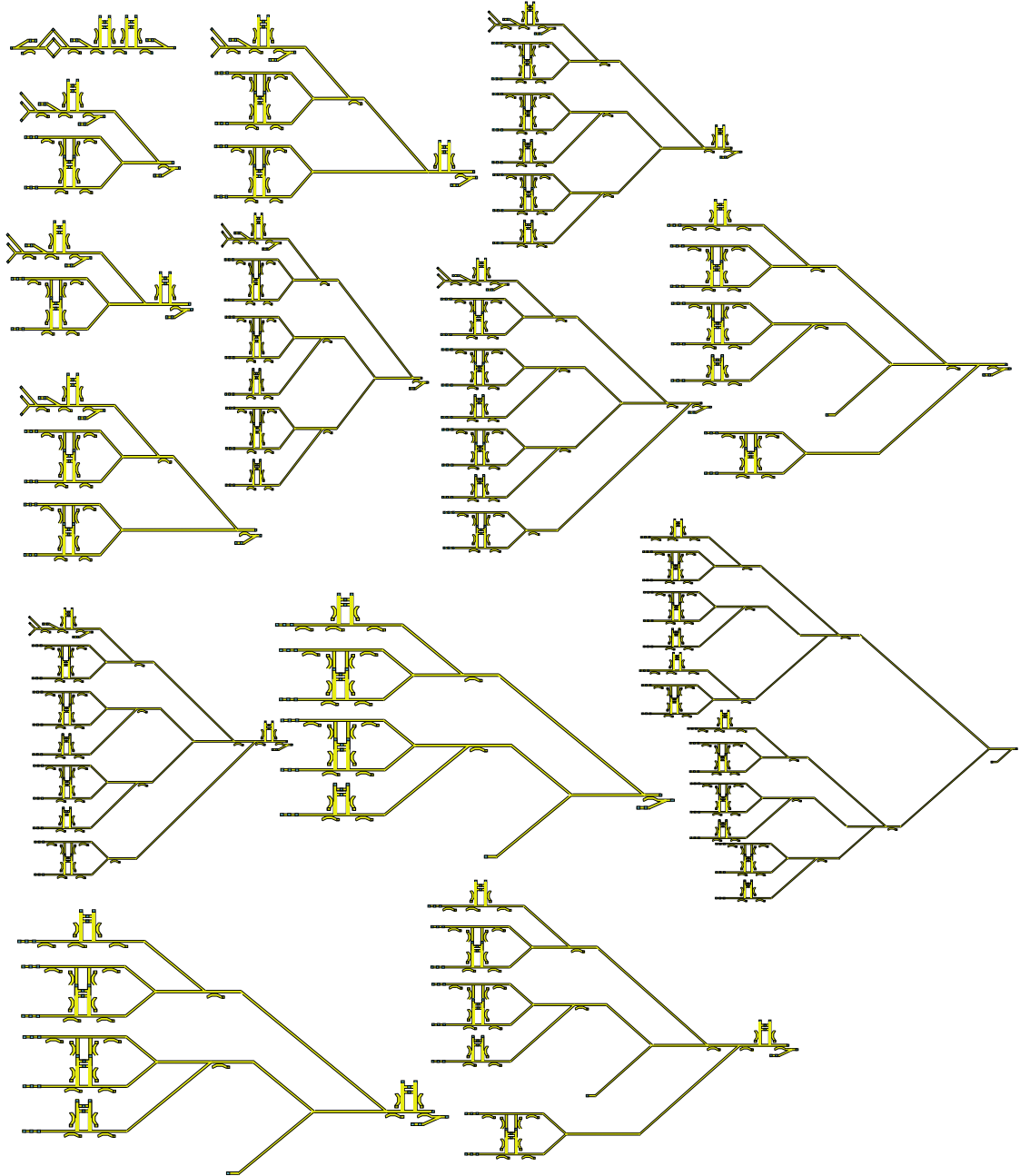}
  \caption{$32$-bit SW Brent Kung Prefix Adder Using Normalizers Only.}
  \label{fig:BKA32_MAJ_S2}
  \vspace{-0.5cm}
\end{figure}

\section{SW Transducer Power Upper Bound}
\label{sec:Transducer Power Upper Bound}

As stated into the introduction, our  goal is to determine the technological limits that need to be reached in order to unleash the SW computing paradigm full potential such that magnonic circuits can outperform CMOS counterparts in terms of energy consumption. In view of the zero power SWs propagation through ferromagnetic waveguides, the overall  magnionic circuit power consumption is determined by the one associated to SWs generation and sensing by means of transducers. Thus we focus our analysis on determining  transducer power consumption acceptable upper bounds that need to be achieve in future transducer  implementations, e.g., ME cells. For this study, we choose a $32$-bit Brent-Kung prefix adder (BKA) \cite{BKA32}, as discussion vehicle and compute the maximum transducer power values that potentially enable a BKA SW implementation able to outperform its \SI{7}{nm} CMOS counterpart. We note that the Brent-Kung adder is a Parallel Prefix Adder (PPA) form of the Carry-Look Ahead adder (CLA) that exhibits structure regularity, low wiring congestion, and reasonable area-performance ratio, which make it quite attractive for practical implementations \cite{BKA32}.  To asses the representativity of our choice, we also determined transducer power upper bound values for:  $32$-bit Wallace Tree Multiplier, $32$-bit Dadda Tree Multiplier, $32$-bit Brent Kung Adder, $64$-bit Dadda Tree Multiplier, $4$-operand $64$-bit Han-Carlson adder, $4$-operand $64$-bit Carry Skip Adder, $32$-bit Multiply Accumulate, $32$-bit Divider, $17$-bit Galois-Field Multiplier, and $32$-bit Cyclic redundancy check. Our results indicate that Brent Kung Adder requires the lowest transducer upper bound (worst case), therefore, our choice as discussion vehicle is relevant for the purpose of this analysis.

\begin{figure}[t]
\centering
  \includegraphics[width=0.8\linewidth]{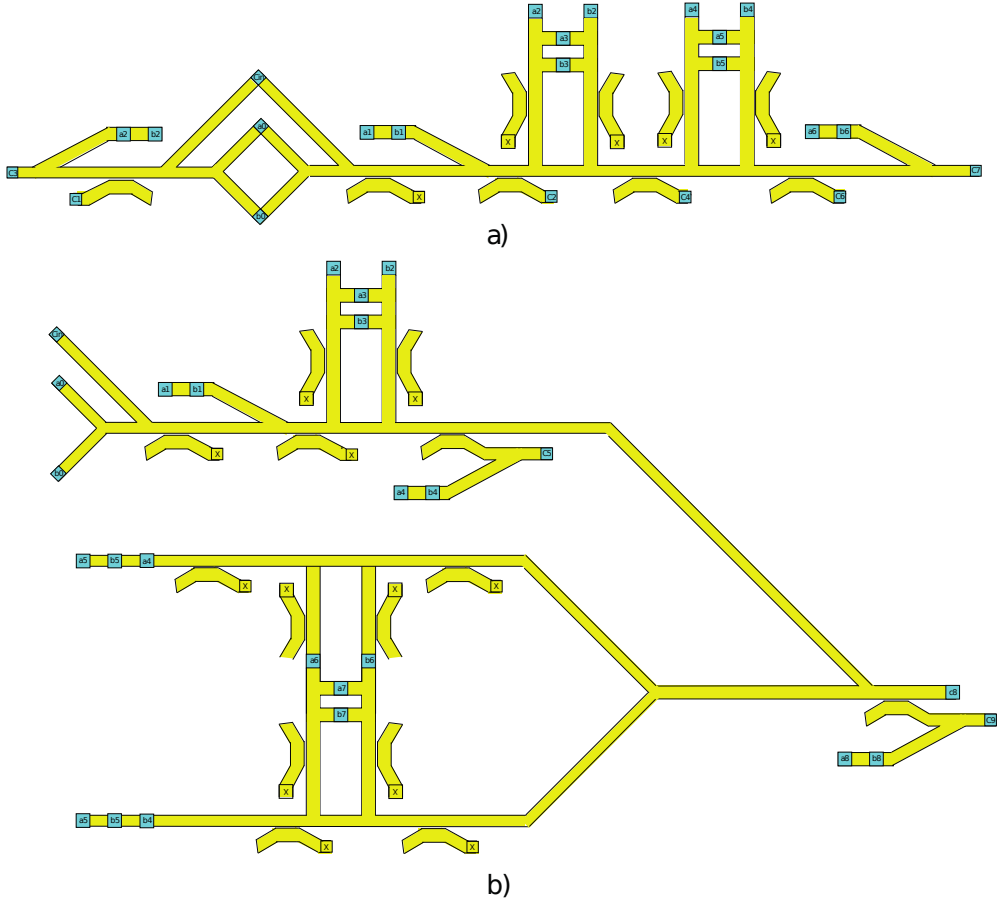}
  \caption{a) Carry-out1 to Carry-out7 Calculation using Normalizers only, b) Carry-out8 to Carry-out9 Calculation using Normalizers only.}
  \label{fig:C1-C9}
  \vspace{-0.5cm}
\end{figure}

\subsection{Possible Implementations}
We evaluate different $32$-bit BKA SW implementations based on Majority gates and compare them with the \SI{7}{nm} CMOS implementation. Note that all our SW circuits rely on the majority tailored implementation method introduced in \cite{BKA8} for the $8$-bit BKA case. As previously mentioned SW gate cascading is not straightforward \cite{amahmoud1} and to this end we evaluated $32$-bit BKA implementations built with: (i) Ideal gate cascading (S1), (ii) Normalizers after each logic gate (S2), (iii) Normalizers and signal conversion back and forth between the electrical and spin wave domain (S3), and (iv) All-in-SW approach (S4). Note that in the implementations, we utilized a combination of fanout enabled ladder shaped Majority gate, programmable logic gates  \cite{fanout, fanout10,fanout11}, triangle shape Majority gate  \cite{mahmoud2021fanout}, in-line Majority gate \cite{inline}, and normalizers (directional couplers)  \cite{amahmoud1}. 

Regardless of the gate cascading method the $32$-bit BKA requires $98$ transducers as it has $65$ inputs and $33$ outputs. This is the case for S1, which assumes direct SW gate cascading, i.e., no normalizers or signal conversion between electrical and SW domain are required to build the adder, and provides the best possible but practically unachievable adder performance.  S2, which provides practically achievable performance data, makes use of directional couplers to normalize SW gate outputs. Figure \ref{fig:BKA32_MAJ_S2} presents its structure, which contains $1040$ transducers as a result of gate replication induced by: 1) unavailability of SW gates with larger than $4$ fanout, 2) unavailability of SW splitters and amplifiers, and 3) layout limitations (waveguides crossovers are not allowed). For example, Figure \ref{fig:C1-C9} presents the SW circuit for calculating the carry-outs C1, C2, C3, \ldots, C9. As it can be observed from the Figure, C1 to C7 are calculated using the SW circuit in Figure \ref{fig:C1-C9} a), which requires $17$ excitation transducers. On the other hand, C8 to C9 are detected using the SW circuit in Figure \ref{fig:C1-C9} b), which requires $23$ excitation transducers, where $9$ transducers are replicated because of fanout limitations.  S3  diminishes the number of required replication and Figure \ref{fig:C1-C9} presents the SW circuit for calculating C1, C2, C3, \ldots, C9 by utilising normalizers and domain conversion (SW to/from electrical). This implementation requires a total of $43$ transducers including excitation, intermediate, and detection transducers, whereas S2 implementation requires $49$, thus we save $6$ transducers for the calculation of the first $9$ carry-outs.

\begin{figure}[t]
\centering
  \includegraphics[width=0.8\linewidth]{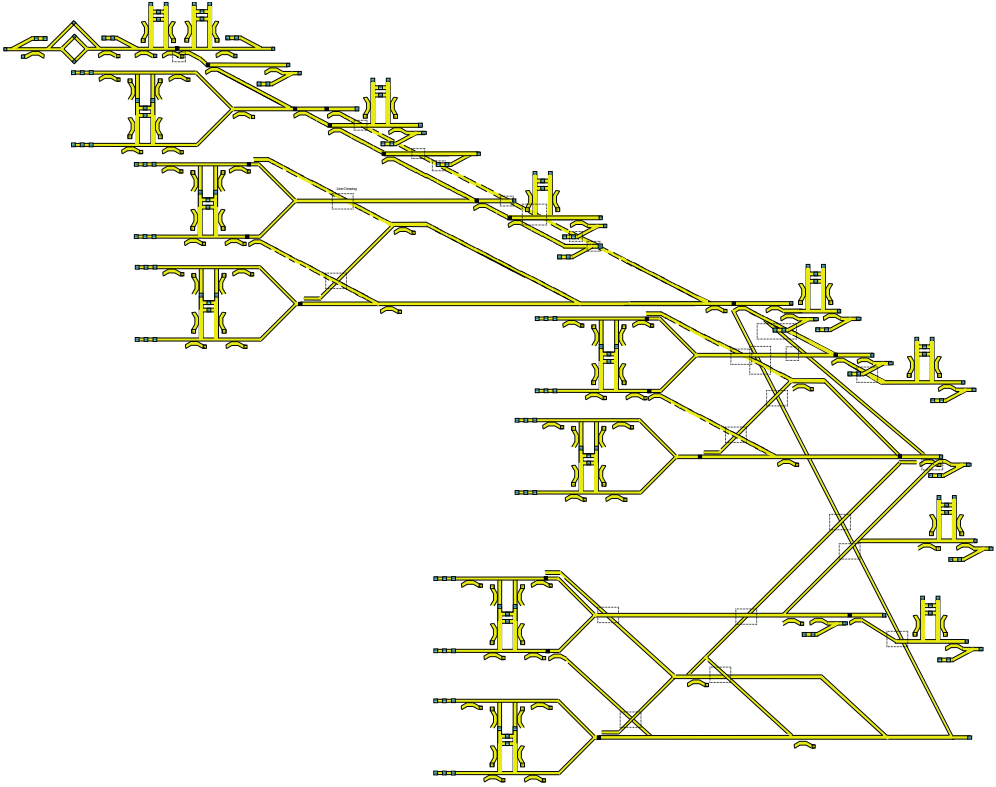}
  \caption{$32$-bit SW Brent Kung Prefix Adder Using Hybrid Approach.}
  \label{fig:BKA32_MAJ_S4}
  \vspace{-0.6cm}
\end{figure}

However, as back and forth domain conversion cost is not yet available, the actual advantage of S3 cannot be accurately  assessed. S4 implementation makes use of normalizers, splitters, amplifiers, and enables line crossover and its structure depicted in  Figure \ref{fig:BKA32_MAJ_S4} makes use of $98$ transducers ($65$ excitation and $33$ detection transducers), $72$ splitters (directional couplers), and $72$ amplifiers.

\begin{figure}[t]
\centering
  \includegraphics[width=0.8\linewidth]{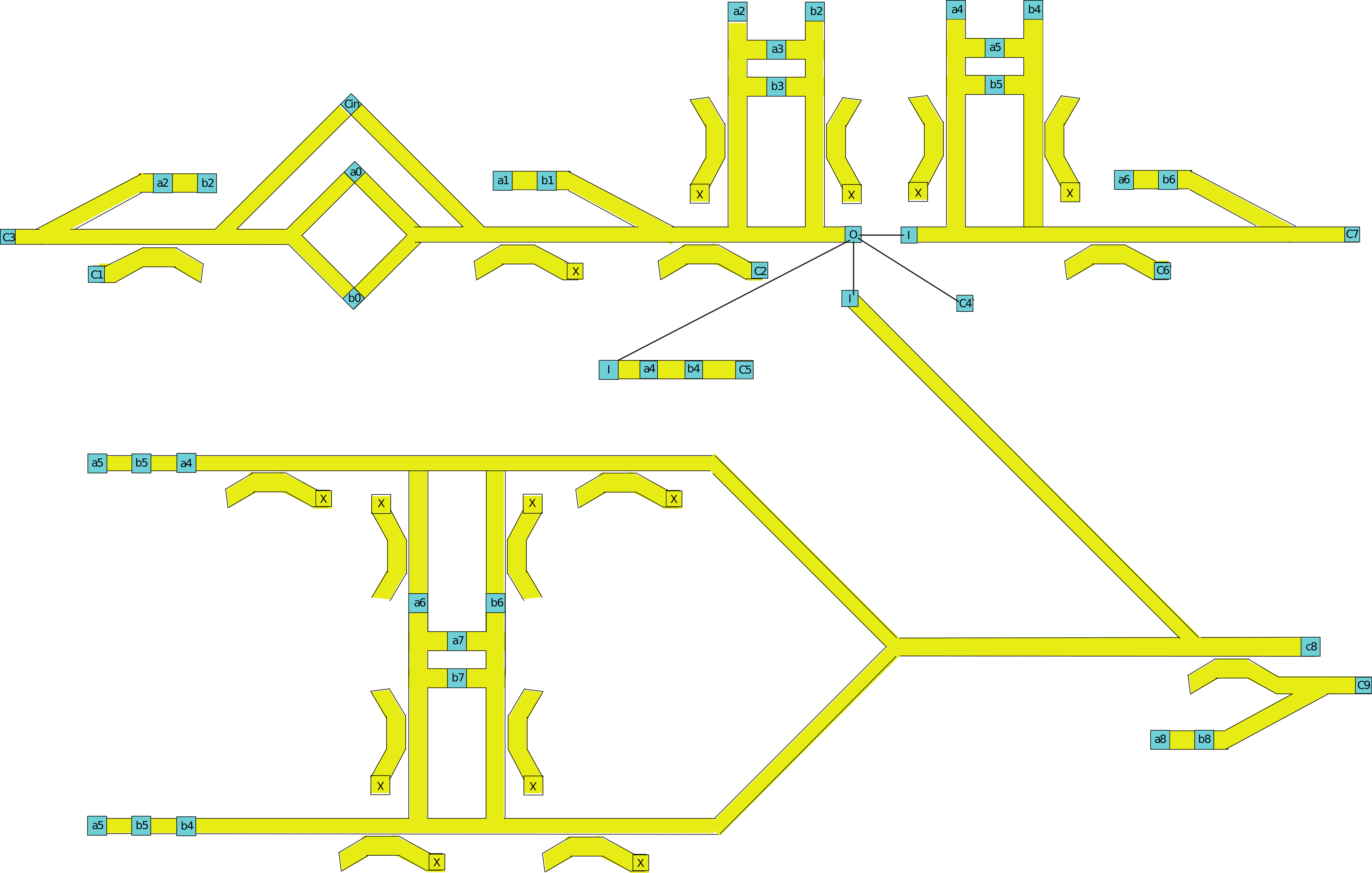}
  \caption{Carry-out1 to Carry-out9 Calculation using Normalizers and Converters.}
  \label{fig:C1-C9b}
  \vspace{-0.6cm}
\end{figure}

\subsection{Transducer Power Upper Bound}

To determine the transducer power consumption upper bound we first need to estimate the power and delay of our reference, i.e.,   CMOS $32$-bit BKA. For this we utilize a commercial \SI{7}{nm} FinFET technology, with regular threshold voltage standard cells, and typical process corner ($V_{DD}$=\SI{0.7}{V}, T=$25^{\circ}C$). The adder was evaluated by means of Cadence simulation, which reported a power consumption of \SI{2.58}{\mu W} and a delay of \SI{1.033}{ns} that translates to an energy consumption of approximately \SI{2.67}{fJ} for the \SI{7}{nm} CMOS $32$-bit BKA. In order to outperform the \SI{7}{nm} CMOS BKA, the SW implementations should exhibit a maximum energy consumption $E_{SW} < E_{CMOS}$, and based on this, we can determine the performance constraint that the transducer needs to fulfil.

To evaluate the delay of a spin wave implementation, we have to identify its critical path, evaluate its physical length and determine the number of transducers it contains. Considering the ladder shape Majority (and their programmable logic gate version) gates \cite{fanout, fanout10,fanout11} and assuming that the maximum propagation length per Majority gate is \SI{336}{nm}, we can evaluate the length of the input SWs trajectory towards outputs in each implementation. First, we note that the critical path encompasses $16$ Majority gates for S1, S2, S3, and S4. Based on this we derive the following critical path lengths:  (i) \SI{5.4}{\mu m} for S1, (ii) \SI{50}{\mu m} for S2, (iii) \SI{43}{\mu m} for S3, and (iv) \SI{85}{\mu m} for S4. Although S3 has the shortest critical path length because it includes the least amount of directional couplers, it does not have the shortest delay because of the domain conversion circuitry. S4 has the longest critical path length because it make use of  amplifiers and splitters to avoid  transducer replications.

To derive the actual critical path delay, the SW propagation speed is required, which equals the SW group velocity that can be  obtained from the dispersion relation material specific slope. Based on the critical path length and SW group velocity, we calculated the delay of the $4$ implementations based on CoFeB waveguide as this material provides the highest SW group velocity. In addition, the following assumptions were made for the delay of the separate elements: \SI{0.42}{ns} transducer delay \cite{Excitation_table_ref16}, a \SI{20}{ns} normalizer delay \cite{amahmoud1}, a \SI{0.03}{ns} peripheral circuit for the converters \cite{Excitation_table_ref16}. Based on this, we derive the following overall delays: (i) \SI{1.92}{ns} for S1, (ii) \SI{14.3}{ns} for S2, (iii) \SI{20}{ns} for S3, and (iv) \SI{24.3}{ns} for S4.  

To proceed with the investigation on the SW adder energy consumption, we concentrate on power consumption estimation. Assuming zero power SW propagation through waveguides as SW doesn't require electron movement and just electron spinning, we can estimate the energy consumption as $E_\mathrm{SW} = TN \times PT \times \text{Delay}$, where $TN$ is the number of transducers in the implementation, $PT$ the power consumed by one transducer, and $\text{Delay}$ the time necessary to excite a SW. Given that in order to outperform CMOS $E_{SW} < E_{CMOS}$, the transducer power consumption upper bound can be determined as $PT = E_{CMOS}/(TN \times Delay)$. $TN$ is determined by circuit topology and for each design we account one per primary input, one per primary output, and (if the case) the appropriate number of repeaters or converters necessary to interconnect the gates forming the prefix adder circuit, which results in $98$, $1040$, $262$, and $203$ transducers for S1, S2, S3, and S4, respectively. It was assumed that each amplifier consumes $\sqrt{n}$, where $n$ is the amplification level. 

The actual $E_{SW}$  value is dependent on the SW operation mode, which defines the $Delay$ value in its evaluation expression. In Continuous Mode Operation (CMO) \cite{Mode_operation} the transducers are active as long as the SWs are propagating through the circuit, i.e., from SW excitation till the output detection. This means that $Delay$ equals the overall adder delay, i.e., \SI{20}{ns}, \SI{12.3}{ns}, and \SI{24.3}{ns} for S2, S3, and S4, respectively. In Pulse Mode Operation (PMO) \cite{Mode_operation}, transducers are active only for a very short period of time required to initiate their output, which we assume to be \SI{0.42}{ns} for all the implementations. Based on this reasoning, we determined the maximum allowable transducer power consumption $PT$ for the CoFeB implementations under CMO and PMO scenarios as presented in Table \ref{table:1}. 

As one can observe in the Table, CMO puts a high pressure on the transducer performance whereas PMO relaxes it by $1$-$2$ orders of magnitude. Moreover, regardless of the operation mode, the hybrid-based implementation is the most energy effective and allows for the highest $PT$ value. Therefore, our preliminary evaluation indicates that the hybrid-based pulse mode operation approach potentially allows spin wave technology to outperform \SI{7}{nm} CMOS, assuming that transducers with maximal \SI{31}{nW} power consumption are achievable. 

\begin{table}[t]
\caption{Transducer Power Upper Bound.}
\label{table:1}
\centering
  \begin{tabular}{|c|c|c|}
    \hline
    Implementation & \multicolumn{2}{c|}{Maximum Power (nW)} \tabularnewline
    \hline
     & Continuous Mode Operation (CMO) & Pulse Mode Operation (PMO) \tabularnewline
    \hline
    Ideal Case & $17$ & $64.9$ \tabularnewline
    \hline
    Normalizer & $0.18$ & $6.1$ \tabularnewline
    \hline
    Normalizer and Converter & $0.51$ & $24$ \tabularnewline
    \hline
    Hybrid & $0.54$ & $31$ \tabularnewline
    \hline
  \end{tabular}
  \vspace{-0.4cm}
\end{table}

\section{Conclusions}
\label{sec:Conclusion}
In this paper, we assessed Magnonic circuits potential to outperform functionally equivalent CMOS counterparts  in terms of energy consumption. We based our analysis on the fact that SW circuits energy consumption is determined by the energy spent by transducers to generate the input SWs and sense the output SWs, as SWs propagation through ferromagnetic waveguides do not consume noticeable energy. While it has been suggested that magneto-electric (ME) cells would be capable to excite and detect SWs while consuming ultra-low power, they have not been experimentally demonstrated and no figures of merit are available. Thus instead of performing a traditional benchmarking we carried on a reverse engineering investigation in an attempt de determine the ME power consumption upper bound that still make Magnonic circuits outperform CMOS counterparts.  To this end, we assumed a $32$-bit Brent-Kung prefix adder as discussion vehicle and determined the maximum transducer power consumption that still make the SW implementation outperform its \SI{7}{nm} CMOS counterpart. We evaluated different SW implementations that rely on conversion- or normalization-based gate cascading and under continuous or pulse SW generation scenarios. Our evaluations indicated that \SI{31}{nW} is the maximum transducer power consumption for which the $32$-bit Brent-Kung SW implementation outperforms its \SI{7}{nm} CMOS counterpart in term of energy.

\section*{Acknowledgement}

This project has received funding from the European Union's Horizon 2020 research and innovation program under grant agreement No. 801055 "Spin Wave Computing for Ultimately-Scaled Hybrid Low-Power Electronics" – CHIRON. It has also been partially supported by imec’s industrial affiliate program on beyond-CMOS logic. F.V. acknowledges financial support from the Research Foundation–-Flanders (FWO) through grant No.~1S05719N.

\bibliography{references}

\end{document}